\documentclass[twocolumn,showpacs,preprintnumbers]{revtex4}
\usepackage{graphicx}
\usepackage{dcolumn}
\usepackage{bm}

\input{tcilatex}

\begin{document}


\title{ Electronic-structure theory of crystalline insulators under
homogeneous electric field }
\author{Zhi-Rong Liu and Jian Wu }
\affiliation{Center for Advanced Study, Tsinghua University, Beijing 100084, 
People's Republic of China}
\author{Wenhui Duan }
\affiliation{Department of Physics, Tsinghua University, Beijing 100084, 
People's Republic of China}
\date{\today }

\begin{abstract}
Based on the conventional energy band theory, an approach is presented to
describe the electronic structure of crystalline insulators in the presence
of a finite homogeneous electric field. 
The expression of polarization is derived which extends the ``Berry-phase'' 
theory of polarization to nonzero fields. The characteristics of the 
solutions are studied in details and the associations with the existing 
perturbation theories and computational schemes are discussed.

\end{abstract}

\pacs{71.20.-b, 71.15.-m, 77.22.Ej }
\maketitle







In recent years, much attention was paid to the theory of macroscopic
polarization and the effect of homogeneous electric field in crystalline
solids.\cite{1,2,3} Although the energy band theory and the
density-functional theory (DFT) have provided a solid basis in understanding
the equilibrium properties of crystals, the problem of crystalline
insulators in the presence of a homogeneous macroscopic electric field was
not fully solved.\cite{3,4} The obstacle comes from the fact that the
electric potential $e\mathbf{\cal E}\cdot \mathbf{r}$ 
is a linear term in the spatial coordinates, which
violates the periodicity condition underlying Bloch's theorem. Moreover,
strictly speaking, there is no ground state for a crystalline system under
an external field. On the other hand, crystalline insulators are
experimentally stable when the applied field is not too strong, and the
field-dependent response of the systems (such as dielectric and
piezoelectric properties) is well defined in the measurements. This looks
like a paradox. It appears that some works are needed to bridge the gap
between the theory and the experiment. Some progress within the framework of
perturbation theory has been achieved in this direction,\cite{4,5,6,7}
nevertheless finite electric fields can not be handled directly in 
such theories due to the
weakness of the perturbation approach. Computational schemes applicable 
at finite electric fields were developed.\cite{8,9,10} 
In such schemes, the non-periodic 
electric potential term $e\mathbf{\cal E}\cdot \langle \mathbf{r} \rangle$ 
in the energy functional is replaced by $-\mathbf{{\cal E}\cdot P}$ where the 
polarization $\mathbf{P}$ is expressed in the Berry-Phase approach 
that holds the periodicity condition. They are effective in numerical 
computing, but the physical basis is not very clear.


On the basis of a band-like electronic-structure theory 
under finite electric fields, this paper aims at providing 
basic theoretical understandings on the subject.
The expression of macroscopic polarization is derived to 
extend the ``Berry-phase'' theory of polarization into the cases of 
nonzero fields. It is shown that the formalism can be transferred 
into a conventional eigen problem in $(\mathbf{r,k})$-space.
The properties of the solutions are identified, 
and the connections with the existing perturbation theories and 
computational schemes are discussed.

\begin{figure}[]
\includegraphics[width=8.5cm]{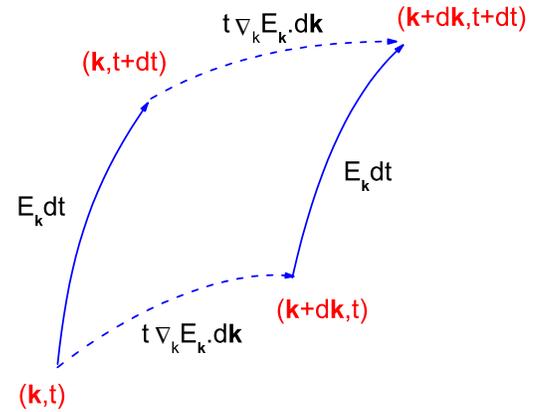}
\caption{Schematic graphics of the change of the phase $E_{\mathbf{k}}t$
when the system evolves from the state ($\mathbf{k}$, $t$) to ($\mathbf{k}+d%
\mathbf{k}$, $t+dt$). If we assume the phase keeps unchanged when the system
transit from $\mathbf{k}$ to $\mathbf{k}+d\mathbf{k}$ (process of dash
line), the total variation of phase is $E_{\mathbf{k}}dt$. }
\label{fig01}
\end{figure}

The theory can be developed as what were done in Refs.~5 and 11, or by 
use of the technique of the crystal momentum representation\cite{12}. 
Here we present a simplified derivation that keeps the insights in physics.

Our approach is based on the conventional energy band theory. According to
the Bloch's theorem, the eigen-state of an electron moving in a periodic
potential $V(\mathbf{r})$ can be expressed as 
\begin{equation}
\psi _{\mathbf{k}}(\mathbf{r},t)=e^{-iE_{\mathbf{k}}t/\hbar }e^{i\mathbf{%
k\cdot r}} u_{\mathbf{k}}(\mathbf{r}),
\end{equation}
where $u_{\mathbf{k}}(\mathbf{r})$ is a periodic function. When there exists
an external electric field $\mathbf{\cal E}$, the Hamiltonian of the system is
given as: 
\begin{equation}
H=-\frac{\hbar ^{2}}{2m}\nabla ^{2}+V(\mathbf{r})+e\mathbf{\cal E}\cdot \mathbf{r}%
.
\end{equation}
In this case, $\mathbf{k}$ is no longer the good quantum number to specify
the eigen-state. In the general solid state physics,\cite{11} the response
of solids to an electric field is described as the constant-velocity
movement of the wave vector $\mathbf{k}$, i.e., 
\begin{equation}
\hbar \frac{d\mathbf{k}}{dt}=-e\mathbf{\cal E}.
\end{equation}%
The bands are either fully filled or empty in insulators, so a static
equilibrium state is reached regardless of the movement of $\mathbf{k}$.
Since $\mathbf{k}$ varies with the time $t$, the differential of Eq.~(1)
with respect to $t$ yields: 
\begin{equation}
i\hbar \frac{\partial }{\partial t}\psi _{\mathbf{k}}=e^{-iE_{\mathbf{k}%
}t/\hbar }e^{i\mathbf{k\cdot r}}\left( E_{\mathbf{k}}-\hbar \frac{d\mathbf{k}%
}{dt}\cdot \mathbf{r}+i\hbar \frac{d\mathbf{k}}{dt}\cdot \nabla _{\mathbf{k}%
}\right) u_{\mathbf{k}}(\mathbf{r}).
\end{equation}%
A key point in deriving the above equation is that we assume the phase $E_{%
\mathbf{k}}t$ keeps unchanged when the system transits between different $%
\mathbf{k}$ states, so $\nabla _{\mathbf{k}}E_{\mathbf{k}}$ is ignored in
the analysis (see Fig.~1). It can be demonstrated more clearly as
follows: if the system transits from an initial wave-function $\psi _{%
\mathbf{k}}$ to a final one $\psi _{\mathbf{k\mathrm{^{\prime}}}}$, then for
a second initial wave-function $\psi _{\mathbf{k}}\exp(i\varphi_0)$, the
system will transit into the final one $\psi _{\mathbf{k\mathrm{^{\prime}}}%
}\exp(i\varphi_0)$, i.e., the initial phase keeps unchanged. Such a process
is invariant for time translation. In contrast, if the term $t\nabla_{%
\mathbf{k}}E_{\mathbf{k}}\cdot d\mathbf{k}$ in Fig.~1 (dash line) is
included, the symmetry of the system for time translation will be violated.

Based on the above analysis and the Schr\"{o}dinger's equation $i\hbar {\partial 
}\psi _{\mathbf{k}}/{\partial t}=H\psi _{\mathbf{k}}$, we get the main
equation of crystalline insulators under electric field as: 
\begin{equation}
(H_{0}+ie\mathbf{\cal E}\cdot \nabla _{\mathbf{k}})u_{\mathbf{k}}=E_{\mathbf{k}%
}u_{\mathbf{k}}(\mathbf{r}),
\end{equation}%
where  
\begin{equation}
H_{0}=-\frac{\hbar ^{2}}{2m}\nabla ^{2}-i\frac{\hbar ^{2}}{m}\mathbf{k}\cdot
\nabla +\frac{\hbar ^{2}\mathbf{k}^{2}}{2m}+V(\mathbf{r}).
\end{equation}%
It can be seen that the introduction of an electric field causes
interactions between different $\mathbf{k}$ states, and $u_{\mathbf{k}}(%
\mathbf{r})$ can be no longer solved independently. However, the periodicity
condition in $\mathbf{r}$-space can be safely applied upon Eq.~(5), while
Eq.~(2) is not periodic, thus it overcomes the difficulties mentioned above. 
Essentially, Eq.~(5) is equivalent to what is obtained in Refs.~5 and 11, 
but it is more convenient for the following analyzing.

To one's surprise, the calculation of polarization in electronic structure 
theory remained a challenge until the Berry-phase theory of polarization 
was developed in the early 1990s.\cite{1,2,13} Although such 
polarization theory has been successfully applied in many cases, 
its validity under 
nonzero electric field was not proven in any publications. 
With Eq.~(5), the expression of polarization at finite fields 
can be derived easily\cite{14} .

Consider an infinitesimal variation of electric field, $\delta \mathbf{\cal E}$,
in the system. The linear terms of Eq.~(5) with respect to $\delta \mathbf{\cal E}
$ yields 
\begin{equation}
(H_{0}+ie\mathbf{\cal E}\cdot \nabla _{\mathbf{k}})\frac{\delta u_{\mathbf{k}}}{%
\delta \mathbf{\cal E}}+ie\nabla _{\mathbf{k}}u_{\mathbf{k}}=E_{\mathbf{k}}\frac{%
\delta u_{\mathbf{k}}}{\delta \mathbf{\cal E}}+\frac{\delta E_{\mathbf{k}}}{%
\delta \mathbf{\cal E}}u_{\mathbf{k}}.
\end{equation}%
Multiplying the above equation by $u_{\mathbf{k}}^{\ast }$ and integrating it,
one gets 
\begin{equation}
\dint \frac{\delta E_{\mathbf{k}}}{\delta \mathbf{\cal E}}d\mathbf{k}=\dint
ie\langle u_{\mathbf{k}}|\nabla _{\mathbf{k}}|u_{\mathbf{k}}\rangle d\mathbf{%
k}
\end{equation}%
if the periodic condition is adopted in $\mathbf{k}$-space as 
\begin{equation}
u_{\mathbf{k+K}}(\mathbf{r})=e^{-i\mathbf{K\cdot r}}u_{\mathbf{k}}(\mathbf{r}%
),
\end{equation}%
where $\mathbf{K}$ is any reciprocal lattice vector. So the macroscopic
polarization is achieved as 
\begin{eqnarray}
\mathbf{P} &=&-\frac{\delta \left( \frac{1}{(2\pi )^{3}}\dint E_{\mathbf{k}}d%
\mathbf{k}\right) }{\delta \mathbf{\cal E}}  \nonumber \\
&=&-\frac{ie}{(2\pi )^{3}}\dint \langle u_{\mathbf{k}}|\nabla _{\mathbf{k}%
}|u_{\mathbf{k}}\rangle d\mathbf{k},
\end{eqnarray}%
which is consistent with the result given by the Berry-phase theory.\cite%
{1,2} Although the original derivation of the polarization's formula in the 
Berry-phase approach assumed $\mathbf{\cal E}=0$, the above analysis showed that 
the same formula keeps valid for $\mathbf{\cal E}\neq 0$.  

Gauge transformation plays an important role in Eq.~(5). It is noted that
Eq.~(5) is invariant under the following transformation: 
\begin{equation}
\left\{ 
\begin{array}{l}
\widetilde{u}_{\mathbf{k}}=u_{\mathbf{k}}\exp \left[ i\beta (\mathbf{k)}%
\right], \\ 
\widetilde{E}_{\mathbf{k}}=E_{\mathbf{k}}-e\mathbf{\cal E} \cdot
\nabla _{\mathbf{k}} \beta (\mathbf{k}),%
\end{array}%
\right.
\end{equation}
where $\beta (\mathbf{k})$ is an arbitrary function. Such phase 
arbitrariness can not be removed even if the periodic boundary 
condition [Eq.~(9)] is adopted.
This reveals that the polarization in Eq.~(10) is a multivalued 
quantity. 

Gauge transformation will not change the physical phase of the system. 
Within an infinitesimal time interval $dt$, the extra phase of 
$u_{\mathbf{k}}$ cased by gauge transformation is 
$\delta\varphi_1=\nabla _{\mathbf{k}} \beta (\mathbf{k}) \cdot d \mathbf{k}$ 
while that of $e^{-iE_{\mathbf{k}}t/\hbar}$ is 
$\delta\varphi_2=e\mathbf{\cal E} \cdot
\nabla _{\mathbf{k}} \beta (\mathbf{k}) dt/\hbar$. With (3) we arrive 
$\delta\varphi_1+\delta\varphi_2=0$. So the gauge transformation 
just moves a part of the phase from $u_{\mathbf{k}}$ to 
$e^{-iE_{\mathbf{k}}t/\hbar}$ while keep the total phase of 
$u_{\mathbf{k}}e^{-iE_{\mathbf{k}}t/\hbar}$ unchanged. It does not cause 
any measurable effects. Furthermore, such an arbitrariness can be utilized to 
simplify some analyses. By choosing $\beta (\mathbf{k})$ appropriately, 
$E_{\mathbf{k}}$ 
can be transformed into a constant independent of $\mathbf{k}$, i.e., 
$E_{\mathbf{k}}\equiv E$, and thus 
Eq.~(5) becomes a conventional eigen-problem in ($\mathbf{r,k}$)-space:
\begin{equation}
(H_{0}+ie\mathbf{\cal E}\cdot \nabla _{\mathbf{k}})u_{\mathbf{k}}=E u_{\mathbf{k}}(\mathbf{r}),
\end{equation}%
whose properties can be well understood based on the existing knowledge 
of eigen problem. Some properties of the solution are listed in the 
following: (1) if there are $N$ bands for $\mathbf{\cal E}=0$, then Eq.~(12) 
has $N$ physically independent solutions for $\mathbf{\cal E}\neq 0$, each 
of which is associated with a group of physically equivalent solutions as 
$\widetilde{u}_{\mathbf{k}}=u_{\mathbf{k}}\exp \left( i\mathbf{k\cdot R}
\right)$ and $\widetilde{E}=E-e\mathbf{\cal E }\cdot \mathbf{R}$ where 
$\mathbf{R}$ is any lattice site; (2) $\nabla _{\mathbf{k}}\langle 
u_{\mathbf{k}}|u_{\mathbf{k}}\rangle =0$ so that $u_{\mathbf{k}}$ 
can be normalized as $\langle u_{\mathbf{k}}|u_{\mathbf{k}}\rangle =1$, 
which is not self-evident for $\mathbf{\cal E}\neq 0$ 
because different $\mathbf{k}$ states are correlated by Eqs~(5) or (12);
(3) the wavefunctions coming from different bands and with the same 
$\mathbf{k}$ value
are orthogonal, i.e., $ \langle u_{\mathbf{k}}^\alpha|u_{\mathbf{k}}^\beta\rangle =0$ ,
where $\alpha$ and $\beta$ denote different bands.

\begin{figure}[tbp]
\includegraphics[width=8.5cm]{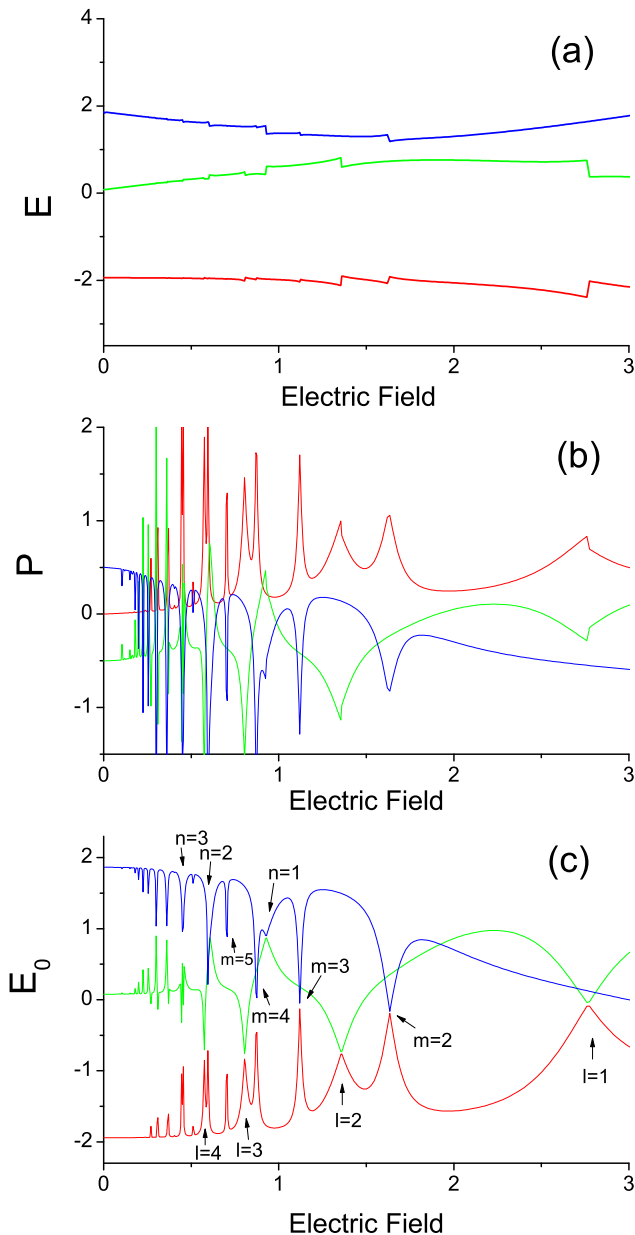}
\caption{ (a) Band energy $E$, (b) polarization $P$ and (c) usual energy 
with polarization's contribution excluded, $E_0=E+\mathbf{{\cal E}\cdot P}$, 
of three bands for the tigh-binding model (see Eq.~(14)) as functions of 
electric field. Labels $l$, $m$ and $n$ in Fig.~2(c) are used to specify 
the index $n$ of peaks between different bands (see Eq.~(16)).
 }
\label{fig02}
\end{figure}

Eq.~(12) is equivalent to the variational analysis on the energy functional:
\begin{equation}
\langle E \rangle =\frac{1}{(2\pi)^3}\dint
\langle u_{\mathbf{k}}|H_0|u_{\mathbf{k}}\rangle d\mathbf{k}
-\mathbf{{\cal E}\cdot P}
\end{equation}%
with $\mathbf{P}$ in Eq.~(10), which is the basis of the perturbation 
theories and computational schemes in references\cite{4,9,10}.

To learn more characteristics of the solution, we conduct calculations on 
a 1D tight-binding Hamiltonian with three bands:\cite{8}
\begin{equation}
H=\sum_j \left \{ \epsilon_jc_j^{\dagger}c_j
+t\left[ c_j^{\dagger}c_{j+1}+c_{j+1}^{\dagger}c_j
\right] \right\},
\end{equation}
where the site energy is given as 
$\epsilon_{3m+k}=\Delta\cos(\alpha-2\pi k/3)$. $\alpha$, $t$ and $\Delta$ 
are three parameters of the Hamiltonian. The position operator is set 
to be $x=\sum_j(j/3)c_j^{\dagger}c_j$ since there are three atoms in 
each unit cell.

In Fig.~2, the properties of three bands are reported as functions of 
electric field for the parameter $e=t=\Delta=1$ and $\alpha=0$. 
The most extraordinary feature at finite electric fields is that there 
are many peaks in the curves of polarization $P$ and the usual energy 
with polarization's contribution excluded, $E_0$ 
($E_0=\langle H_0 \rangle=E+\mathbf{{\cal E}\cdot P}$). 
Correspondingly, the curves of band energy $E$ are discontinuous with many gaps.

The peaks are caused by the overlap of the energy of different bands. 
According to Eq.~(12), the dielectric susceptibility of 
the band $\alpha$ is given as:
\begin{equation}
\chi_\alpha=\left(\frac{a}{2\pi}\right)^3.
\frac{2e^2}{(2\pi)^3 \varepsilon_0}{\rm Re} \sum_{\beta\neq\alpha}
\frac{\langle u^\beta|\nabla _{\mathbf{k}}|u^\alpha\rangle^*
\langle u^\beta|\nabla _{\mathbf{k}}|u^\alpha\rangle}
{E^\beta-E^\alpha}.
\end{equation}
Note that the band energy is a multivalued quantity with a term 
$-ne{\cal E }a$ where $a$ is the crystal 
lattice in the current one-dimensional case. So if the electric 
field satisfies the following relation:
\begin{equation}
{\cal E}=\frac{1}{n}\cdot\frac{E^\beta-E^\alpha}{ea}~~~~ (n=\pm1,\pm2,...),
\end{equation}
the dielectric susceptibility diverges and its sign changes when 
the electric field goes across such points, which results in the 
peaks of the polarization. 

The revealed properties of polarization at finite fields here are 
helpful in understanding the convergency of the perturbation theories. 
The electric filed associated with the peaks of polarization 
is a kind of ``singular point''. From Eq.~(16) it is known that 
there are more and more such singular points when the electric 
filed is getting smaller. Especially, the zero field ${\cal E}=0$ 
is a non-isolated singular point of which there are always 
other singular points within any small neighborhoods. 
Therefore, if the properties of the system are expanded in the 
form of power series in the electric field, the series are not 
convergent within any finite neighborhood of radius centered 
about ${\cal E}=0$. That is the reason why the perturbation 
theories in continuum formulation are valid only for 
infinitesimal fields.\cite{4,5}

When the band energy $E$ has a minimal value, the solution of ground 
state can be obtained 
by minimizing the energy functional Eq.~(13). Unfortunately, the 
minimal value of $E$ does not exist in continuum formulation due to the 
term $-ne{\cal E }a$. A technique to remove such difficulty is to 
perform a discretization on the $\mathbf{k}$-space, i.e., to 
restrict the maximal value of the number $n$, which has been adopted 
in numerical schemes.\cite{9,10} However, when the electric field 
is large enough to shift the higher band to below the lower band, 
i.e., ${\cal E}>(E^\beta-E^\alpha) /(N_kea)$ where $N_k$ is the discretization
number in $\mathbf{k}$-space, the minimizing procedure will fail in 
obtaining the ground state. That is the phenomena of collapse appeared 
in the numerical schemes.\cite{9,10} It is clear that such phenomena 
are associated with the breaking of the numerical schemes, but not 
with the breaking of the system such as Zener tunneling\cite{15}.

\begin{figure}[tbp]
\includegraphics[width=6.5cm]{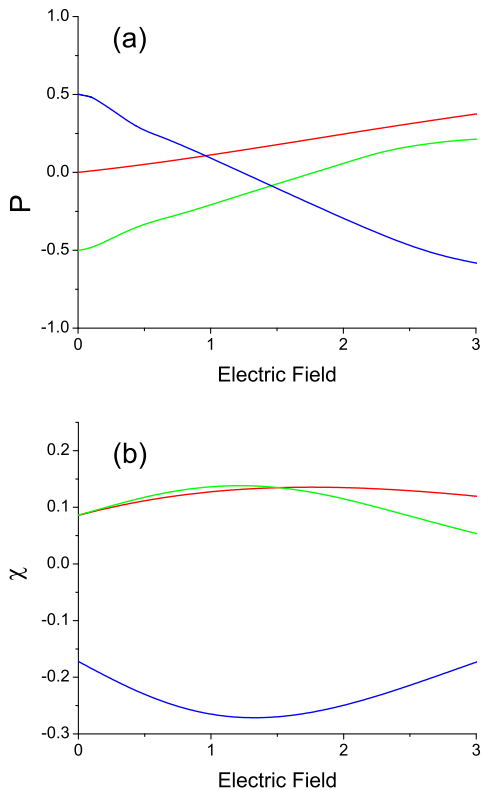}
\caption{ (a) The average polarization and (b) dielectric susceptibility 
of the three-band model. The peaks in Fig.~2 have been smoothed out to 
provide results comparable with experiments.
 }
\label{fig03}
\end{figure}

Although there are infinite peaks in the calculated 
curve of polarization around ${\cal E}=0$, the properties 
of the system is smooth enough in experiments. So it is necessary 
to have a short discussion on the width of the peaks. 
After a simple analysis of degenerate perturbation around the peak, 
it is shown that the width of the peaks, $\Delta {\cal E}$, 
is approximately estimated as:
\begin{equation}
\frac{\Delta {\cal E}}{\cal E}={\cal E}
\left[ \frac{ a^3 \varepsilon_0 \chi}{2(E^\beta_0-E^\alpha_0)}
\right]^{1/2},
\end{equation}
where $E^\alpha_0$ and $E^\beta_0$ are the average band energy 
when there is no electric field. For a numerical estimate, 
we set $\chi=10$, $a=2${\AA} and $E^\beta_0-E^\alpha_0=2{\rm eV}$. 
Then we get $[a^3 \chi/(2(E^\beta_0-E^\alpha_0))]^{1/2}=
3.3\times 10^{-11}$ (in SI unit). It is obvious that such narrow 
peaks at small electric fields are too difficult to be observed 
experimentally. Therefore, the properties obtained in experiments 
reflect an average effect where the peaks are smoothed out.
In Fig.~3 we depict the curves of the average polarization and 
dielectric susceptibility after the peaks are removed. 
Nonlinear effect can be observed. 
The dielectric susceptibility is non-monotonic, 
which means some high-order susceptibility terms are negative.

It is noted that some issues discussed in this paper, such as 
the multivalued property of the polarization and the absence of 
ground state at finite electric fields, are caused by the 
periodic boundary conditions imposed on the insulators. If 
we consider a finite piece of a solid and surface effects, 
such problems will disappear. That is what should be kept in mind 
when one considers the effect of electric fields on insulators.

We would like to thank Dr.~Jian-She Liu for helpful discussions. This
work was supported by the Chinese State Key Program of Basic Research
Development (Grant No.TG2000067108), the National High Technology
Research and Development Program (Grant No.2002AA311153) and 
the National Natural Science Foundation of China.


\appendix{}

\vspace{1cm}

{\bf{Note: the following appendix are provided for the referees' 
reference, but not for publication.}}

\section{rigorous derivation of polarization}

In crystalline insulators, the sole value of polarization is
meaningless since its value depends on the state of the surface. 
The meaningful one, which can be compared with experiments, 
is the change of polarization induced by
external factors such as electric field or stress.  
So a rigorous derivation of polarization is based on the 
analysis on the variation of the system with respect to the 
external electric field.

Denote the expected value of $H_0$ as 
\begin{equation}
E_0=\langle u | H_0 | u \rangle. 
\end{equation}
It is a physical quantity that is well defined in crystalline 
insulators. Then the physical energy of the system can be written as:
\begin{equation}
\langle E \rangle=E_0-\mathbf{P}\cdot \mathbf{\cal E},
\end{equation}
where $\mathbf{P}$ is the physical polarization whose expression is 
unknown at this point. For a ground (stable) state, the energy 
functional reaches its minimum, i.e., 
\begin{equation}
\delta\langle E \rangle=0
\end{equation}
for any infinitesimal variation of wavefunctions. Now, considering the 
variation of wavefunctions in a virtual process 
$\mathbf{\cal E}\rightarrow \mathbf{\cal E}+\delta\mathbf{\cal E}$, 
from Eqs.~(A2) and (A3) we have
\begin{equation}
\delta \langle E \rangle=\frac{d E_0}{d \mathbf{\cal E}} \cdot 
\delta \mathbf{\cal E}-\left(\frac{d\mathbf{P}}{d\mathbf{\cal E}}\cdot
 \delta\mathbf{\cal E}\right)\cdot \mathbf{\cal E}=0.
\end{equation}
So 
\begin{equation}
\frac{d\mathbf{P}}{d\mathbf{\cal E}}\cdot  \mathbf{\cal E}
=\frac{d E_0}{d \mathbf{\cal E}} .
\end{equation}
It relates $\mathbf{P}$ to the well-defined quantity $E_0$.

Now define the Berry-Phase polarization as:
\begin{equation}
\mathbf{P}_{\rm B} =-\frac{ie}{(2\pi )^{3}}\dint \langle u_{\mathbf{k}}|\nabla _{\mathbf{k}}|u_{\mathbf{k}}\rangle d\mathbf{k}.
\end{equation}
It is not a physical quantity since it changes with gauge transformation.
From Eq.~(5), we have
\begin{equation}
\mathbf{\cal E}\cdot\mathbf{P}_{\rm B} 
=E_0-\frac{1}{(2\pi )^{3}}\dint E_{\mathbf{k}} d\mathbf{k}.
\end{equation}
Conducting differential on the above equation with respect to
$\mathbf{\cal E}$, we have
\begin{equation}
\mathbf{P}_{\rm B}+\mathbf{\cal E}\cdot\frac{d\mathbf{P}_{\rm B} }
{d\mathbf{\cal E}}
=\frac{d E_0}{d\mathbf{\cal E}}
-\frac{1}{(2\pi )^{3}} \frac{d\dint E_{\mathbf{k}} d\mathbf{k}}
{d\mathbf{\cal E}}.
\end{equation}
Combined with Eqs.~(10), (A5) and (A6), we reach
\begin{equation}
\frac{d\mathbf{P}_{\rm B} }{d\mathbf{\cal E}}
=\frac{d\mathbf{P} }{d\mathbf{\cal E}}.
\end{equation}
So, the polarization defined in Berry-Phase expression, $\mathbf{P}_{\rm B}$, 
is equal to the physical polarization $\mathbf{P}$ except a constant.

\section{Normalization and orthogonality of $u_{\mathbf{k}}$}

Under finite electric field, there are correlations between 
wavefunctions $u_{\mathbf{k}}$ with different $\mathbf{k}$ values. 
However, it can be shown that $u_{\mathbf{k}}$ are normalized and 
orthogonal as the case of $\mathbf{\cal E}=0$. 

Compared with the Schrodinger's equation $i\hbar {\partial } \psi _{%
\mathbf{k}}/\partial t=H\psi _{\mathbf{k}}$, it is recognized that the role
of $\mathbf{k}$ in
Eq.~(5) is similar to that of $t$. A consequence of this similarity
is the conservation of possibility in the $\mathbf{k}$-space: 
\begin{equation}
\nabla _{\mathbf{k}}\langle u_{\mathbf{k}}|u_{\mathbf{k}}\rangle =0,
\end{equation}
which is a physical requirement for the evolution equation. 
So the wavefunction can be normalized as 
$\langle u_{\mathbf{k}}|u_{\mathbf{k}}\rangle =1$ or
$\langle u_{\mathbf{k}}|u_{\mathbf{k}}\rangle =a^3$. 
The difference
between $\mathbf{k}$ and $t$ exists in the boundary condition. In the
Schrodinger's equation, no boundary condition is required in $t$-space,
while in Eq.~(5) $\mathbf{k}$ and $\mathbf{k+K}$ label the same state (with
a possible phase difference), so that the solutions to Eq.~(5) are discrete
(band structure).

In the conventional band theory without electric field, the wave-functions coming
from different bands are orthogonal, i.e., 
\begin{equation}
\langle u_{\mathbf{k}}^\alpha|u_{\mathbf{k}}^\beta\rangle =0,
\end{equation}
where $\alpha$ and $\beta$ denote different bands. Since Eq.~(12) 
is a conventional eigen-problem, the orthogonality gives
\begin{equation}
\sum_{\mathbf{k}}\langle u_{\mathbf{k}}^\alpha|u_{\mathbf{k}}^\beta\rangle =0
\end{equation}
at finite fields. 
In the first glance, Eq.~(B2) is no longer valid. Nevertheless,
to one's surprise, Eq.~(B2) keeps valid in this case. To prove this point, we
transform $u_{\mathbf{k}}^\beta$ into 
\begin{equation}
\widetilde{u}_{\mathbf{k}}^\beta=u_{\mathbf{k}}^\beta\exp( i\mathbf{k\cdot R}),
\end{equation}
($\mathbf{R}$ can be any lattice vector), which is also an eigen
wavefunction of Eq.~(12), thus 
\begin{equation}
\sum_{\mathbf{k}}\langle {u}_{\mathbf{k}}^\alpha|\widetilde{u}_{\mathbf{k}}%
^\beta\rangle =\sum_{\mathbf{k}}\langle u_{\mathbf{k}}^\alpha|u_{\mathbf{k}}%
^\beta\rangle \exp( i\mathbf{k\cdot R})= 0.
\end{equation}
By applying the knowledge of Fourier's transformation on Eq.~(18), we arrive
at Eq.~(B2) with electric fields.

\section{the width of the peak of polarization}

The width of the peak can be evaluated by considering the degenerate 
perturbation analysis on two bands $u^{\alpha}$ and $u^{\beta}$. 
Expanding the wavefunction on the bands $\alpha$ and $\beta$ of 
$\mathbf{\cal E}=0$, the Hamiltonian becomes:
\begin{equation}
H=\left[ 
\begin{array}{cc}
E_0^\alpha+\mathbf{\cal E}\cdot a^3\mathbf{P}^{\alpha\alpha} &
\mathbf{\cal E}\cdot a^3\mathbf{P}^{\alpha\beta} \\
\mathbf{\cal E}\cdot a^3\mathbf{P}^{\beta\alpha} &
E_0^\beta+\mathbf{\cal E}\cdot a^3\mathbf{P}^{\beta\beta}
\end{array}%
\right],
\end{equation}
where
\begin{equation}
\mathbf{P}^{\mu\nu} =-\frac{ie}{(2\pi )^{3}}\dint \langle u_{\mathbf{k}0}^\mu|\nabla _{\mathbf{k}%
}|u_{\mathbf{k}0}^\nu\rangle d\mathbf{k}~~~~(\mu,\nu=\alpha,\beta).
\end{equation}%
Eq.~(C1) describes the transition between bands $\alpha$ and $\beta$, 
where the transition range (width of the peak) is:
\begin{equation}
\frac{\Delta \mathbf{\cal E}}{\mathbf{\cal E}}
=\frac{\mathbf{P}^{\alpha\beta}}
{\mathbf{P}^{\alpha\alpha}-\mathbf{P}^{\beta\beta}}.
\end{equation}%
The degenerate condition requires that
\begin{equation}
E_0^\alpha+\mathbf{\cal E}\cdot a^3\mathbf{P}^{\alpha\alpha}
=E_0^\beta+\mathbf{\cal E}\cdot a^3\mathbf{P}^{\beta\beta},
\end{equation}%
which gives
\begin{equation}
\mathbf{P}^{\alpha\alpha}-\mathbf{P}^{\beta\beta}
=\frac{E_0^\alpha-E_0^\beta}
{\mathbf{\cal E} a^3}.
\end{equation}%
From Eq.~(15), $\mathbf{P}^{\alpha\beta}$ is approximated as
\begin{equation}
\chi\simeq\frac{2a^3|\mathbf{P}^{\alpha\beta}|^2}
{\varepsilon_0 (E_0^\beta-E_0^\alpha)}.
\end{equation}%
Combining Eqs.~(C3,C5,C6), we arrive
\begin{equation}
\frac{\Delta {\cal E}}{\cal E}={\cal E}
\left[ \frac{ a^3 \varepsilon_0 \chi}{2(E^\beta_0-E^\alpha_0)}
\right]^{1/2}.
\end{equation}

\section{iteration solving of the equation}

In the above main text, it is shown that our theory provides 
the basis for existing computational schemes\cite{9,10}. 
Based on the theory, other computational schemes may also be developed.
The calculation in the main text is conducted by solving the conventional 
eigen problem Eq.~(12) with subroutine ZHEEVX in the program packet 
LAPACK, which is applicable for any finite electric fields but requires 
more computer capacity. Here, we demonstrate a second scheme to solve 
Eq.~(5). 

The computational scheme is based on the following iteration:
\begin{equation}
\left\{
\begin{array}{l}
E_{\mathbf{k}}^{(i)}=\langle u_{\mathbf{k}}^{(i-1)}(\mathbf{r})
|H_0+ie\mathbf{\cal E}\cdot \nabla _{\mathbf{k}} |
u_{\mathbf{k}}^{(i-1)}(\mathbf{r}) \rangle \\
(H_{0}-E_{\mathbf{k}}^{(i)})u_{\mathbf{k}}^{(i)}(\mathbf{r})
=-ie\mathbf{\cal E}\cdot \nabla _{\mathbf{k}}u_{\mathbf{k}}^{(i-1)}(\mathbf{r}) 
\end{array}
\right.
\end{equation}%
$\mathbf{k}$-space is discretized into 99 points ($N_{\mathbf{k}}=99$). 
Iteration stops when the discrepancy is smaller than the 
convergence criterion: $|u_{\mathbf{k}}^{(i)}
-u_{\mathbf{k}}^{(i-1)}|<10^{-4}$. A computation example on the 
three-band model is shown in Fig~4. It can be seen that 
the convergence is obtained at $\mathbf{\cal E}\leq 0.2$ for 
the first band (Fig.~4(a)), which exceeds the critical field 
given by Ref.~9 ($E_{\rm gap}/eaN_k=0.012$) by about one order.

\begin{figure}[tb]
\includegraphics[width=8.5cm]{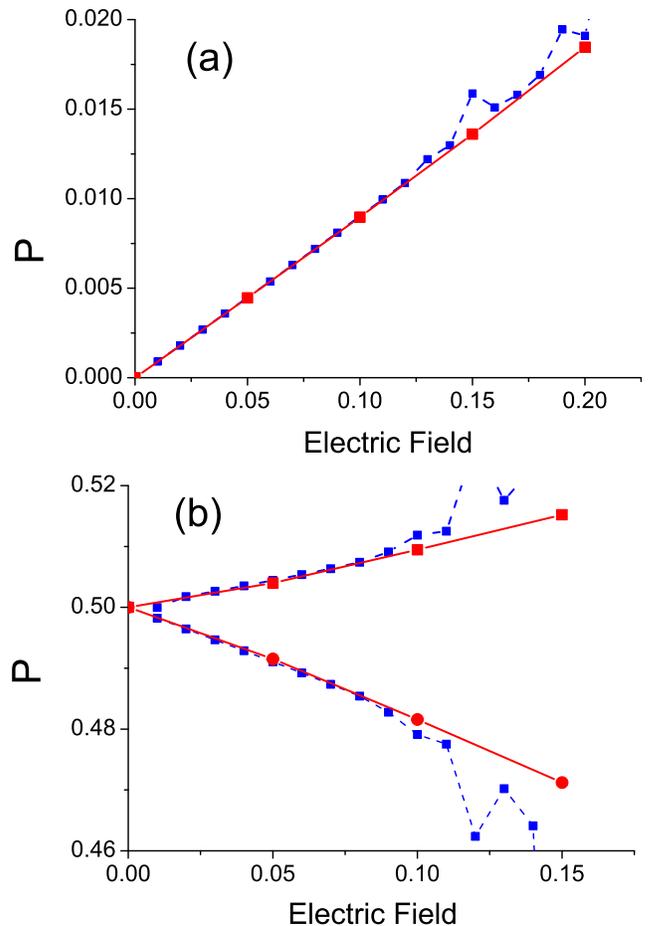}
\caption{ Polarization of three bands given by computational scheme of 
iteration Eq.~(D1) (red), compared with what is given by the 
eigen-problem subroutine of LAPACK (blue). (a) Polarization of the first 
band. (b) Polarization of the second and the third bands.
 }
\label{Appen01}
\end{figure}

\section{Dielectric susceptibility at zero field: a comparison}
The analytic solution for the static susceptibility at zero field 
can be computed from the Kubo formula. In the following we show that 
the susceptibility derived from the theory in this paper is 
consistent with what is given by Kubo formula.

According to the reference (H. Ehrenreich and M. H. Cohen, 
Phys. Rev {\bf 115}, 786), the analytical 
expression of the static 
susceptibility at zero field is given as:
\begin{equation}
\chi=\frac{e^2}{m\pi^2}\sum_{l,l'}  {'}\dint f_0(E_{kl})
f_{l'l}^\mu(\omega_{l'l})^{-2}d^3k,
\end{equation}
where $f_0(E_{kl})$ is the Fermi distribution and
\begin{eqnarray}
f_{l'l}^\mu=\frac{2}{\hbar\omega_{l'l}m}|P_{l'l}^\mu|^2, \\
P_{l'l}^\mu=\frac{1}{\upsilon}\dint u_{kl'}^*P^\mu u_{kl} d^3x, \\
P^{\mu}=-i\hbar \nabla\cdot \hat{\mu}.
\end{eqnarray}
At zero temperature ($T=0$),
\begin{equation}
\chi=\sum_{l}^{\rm occupied} \frac{2e^2\hbar}{m^2\pi^2}
\sum_{l'}  {'} \dint 
\frac{\left| \langle u_{kl'}|\hat{\mu}\cdot \nabla |u_{kl}
\rangle \right|^2}
{\omega_{l'l}^3} d^3k.
\end{equation}

For the current theory, with an infinitesimal variation 
$\delta \mathbf{\cal E}$, the linear term of  Eq.~(5) 
at $\mathbf{\cal E}=0$ is:
\begin{equation}
H_0\frac{\delta u_{\mathbf{k}l}}{\delta \mathbf{\cal E}}
+ie\nabla_{\mathbf{k}} u_{\mathbf{k}l}=
E_{\mathbf{k}l}\frac{\delta u_{\mathbf{k}l}}{\delta \mathbf{\cal E}}
+\frac{\delta E_{\mathbf{k}l}}{\delta \mathbf{\cal E}}u_{\mathbf{k}l}.
\end{equation}
Expand $\frac{\delta u_{\mathbf{k}l}}{\delta \mathbf{\cal E}}$ as
$\frac{\delta u_{\mathbf{k}l}}{\delta \mathbf{\cal E}}
=\sum_{l'}a_{\mathbf{k}l'} u_{\mathbf{k}l'}$, thus
\begin{equation}
\sum_{l'}a_{\mathbf{k}l'} E_{\mathbf{k}l'} u_{\mathbf{k}l'}
+ie\nabla_{\mathbf{k}} u_{\mathbf{k}l}=
\sum_{l'}a_{\mathbf{k}l'} E_{\mathbf{k}l} u_{\mathbf{k}l'}
+\frac{\delta E_{\mathbf{k}l}}{\delta \mathbf{\cal E}}u_{\mathbf{k}l}.
\end{equation}
Integral the above equation with $\langle u_{\mathbf{k}l'} |$ yielding
\begin{equation}
a_{\mathbf{k}l'}=\frac{ie}{\hbar\omega_{l'l}}
\langle u_{\mathbf{k}l'} | \nabla_{\mathbf{k}} u_{\mathbf{k}l }\rangle.
\end{equation}
So the dielectric susceptibility is 
\begin{equation}
\begin{array}{lll}
\frac{\delta \mathbf{P}_l}{\delta \mathbf{\cal E}} &=&
-\frac{ie}{(2\pi)^3}\sum_{l'}\dint \left[
\langle \frac{\delta u_{\mathbf{k}l}}{\delta \mathbf{\cal E}} 
 | \nabla_{\mathbf{k}} u_{\mathbf{k}l }\rangle
+\langle u_{\mathbf{k}l }  
 | \nabla_{\mathbf{k}}
 | \frac{\delta u_{\mathbf{k}l}}{\delta \mathbf{\cal E}}\rangle 
\right]d^3\mathbf{k}  \\
  & = & 
\frac{2e^2}{(2\pi)^3\hbar} {\rm Re} \sum_{l'}
 \dint 
\frac{ 
\langle u_{\mathbf{k}l' } | \nabla_{\mathbf{k}} | u_{\mathbf{k}l} \rangle ^* 
\langle u_{\mathbf{k}l' } | \nabla_{\mathbf{k}} | u_{\mathbf{k}l} \rangle }
{\omega_{l'l}} d^3\mathbf{k}.
\end{array}
\end{equation}

Eq.~(E9) can be further simplified.
Multipling $\nabla_{\mathbf{k}}$ with Eq.~(5) gives:
\begin{equation}
\begin{array}{lll}
& & H_0 \nabla_{\mathbf{k}}u_{\mathbf{k}l} 
-i\frac{\hbar^2}{m}\nabla u_{\mathbf{k}l}
+\frac{\hbar^2\mathbf{k}}{2m} u_{\mathbf{k}l} \\
& = &
E_{\mathbf{k}l}\nabla_{\mathbf{k}} u_{\mathbf{k}l}
+u_{\mathbf{k}l}\nabla_{\mathbf{k}} E_{\mathbf{k}l}.
\end{array}
\end{equation}
Multipling with $\langle u_{\mathbf{k}l'} |$ yields
\begin{equation}
\langle u_{\mathbf{k}l' } | \nabla_{\mathbf{k}} | u_{\mathbf{k}l} \rangle
=\frac{i\hbar}{m}.
\frac{\langle u_{\mathbf{k}l' } | \nabla | u_{\mathbf{k}l} \rangle}
{\omega_{l'l}}.
\end{equation}
So the dielectric susceptibility is 
\begin{equation}
\begin{array}{lll}
\chi & = & \frac{1}{\varepsilon_0}.
\frac{\delta \mathbf{P}_l}{\delta \mathbf{\cal E}} \\
& = &
\frac{2e^2\hbar}{(2\pi)^3\varepsilon_0 m^2} {\rm Re} \sum_{l'}
 \dint 
\frac{ 
\langle u_{\mathbf{k}l' } | \nabla | u_{\mathbf{k}l} \rangle ^* 
\langle u_{\mathbf{k}l' } | \nabla | u_{\mathbf{k}l} \rangle }
{\omega_{l'l}^3} d^3\mathbf{k}.
\end{array}
\end{equation}
Considering the difference of the units used
(SI vs. Gauss), Eqs.~(E12) and (E5) are consistent.

\section{A short note on Zener tunneling}

Essentially, the Zener tunneling is a kinetic effect in which 
the transition rate between different states should be calculated. 
Therefore, Zener tunneling can not be directly solved in the 
framework of this paper or the existing computational schemes\cite{9,10} 
or perturbation theories\cite{5} 
whose purpose is to obtain the eigen states of the system. 
Eigen states are unchanged with time and no transition will occur 
if no extra term is introduced into the Hamiltonian.

Zener tunneling is usually analyzed by semi-classical method. Here 
we illustrate some properties of motion by the theory of this paper.

Firstly, the velocity of the electron is given as:
\begin{equation}
\begin{array}{lll}
\mathbf{\upsilon} & = & \frac{-i\hbar}{m}\dint 
\psi^{*} \nabla \psi d^3\mathbf{r} \\
&= & \frac{\hbar \mathbf{k}}{m}+
\frac{-i\hbar}{m} \langle u_{\mathbf{k} } | \nabla | u_{\mathbf{k}} \rangle .
\end{array}
\end{equation}
Multipling $\nabla_{\mathbf{k}}$ with Eq.~(5) gives:
\begin{equation}
\begin{array}{l}
 H_0 \nabla_{\mathbf{k}}u_{\mathbf{k}} 
+ie\mathbf{\cal E}\cdot \nabla _{\mathbf{k}}\nabla _{\mathbf{k}}u_{\mathbf{k}}
-i\frac{\hbar^2}{m}\nabla u_{\mathbf{k}}
+\frac{\hbar^2\mathbf{k}}{m} u_{\mathbf{k}} \\
=
E_{\mathbf{k}}\nabla_{\mathbf{k}} u_{\mathbf{k}}
+u_{\mathbf{k}}\nabla_{\mathbf{k}} E_{\mathbf{k}}.
\end{array}
\end{equation}
Multipling with $| u_{\mathbf{k}} \rangle$ yields
\begin{equation}
\nabla_{\mathbf{k}} \langle u_{\mathbf{k}} | 
ie\mathbf{\cal E}\cdot \nabla _{\mathbf{k}} | u_{\mathbf{k}} \rangle
-\frac{i\hbar^2}{m} \langle u_{\mathbf{k}} | 
\nabla  | u_{\mathbf{k}} \rangle
+\frac{\hbar^2\mathbf{k}}{m}= \nabla _{\mathbf{k}} E_{\mathbf{k}}.
\end{equation}
So the velocity is
\begin{equation}
\mathbf{\upsilon}=\nabla _{\mathbf{k}} E_{\mathbf{k}}^{0},
\end{equation}
where
\begin{equation}
E_{\mathbf{k}}^{0}=\langle u_{\mathbf{k}} | 
H_0  | u_{\mathbf{k}} \rangle.
\end{equation}

Secondly, we consider the motion of the wave packet. 
Consider a wave packet $(\mathbf{k_0},\mathbf{r_0})$ at $t_0$.
Then at time $t$, the $\mathbf{k}$ will evolve into 
$\mathbf{k}=\mathbf{k_0}-e\mathbf{\cal E}(t-t_0)/\hbar$, and
\begin{equation}
\mathbf{r}-\mathbf{r_0}=\dint_{t_0}^{t} \mathbf{\upsilon}dt
=\dint_{t_0}^{t} \nabla _{\mathbf{k}} E_{\mathbf{k}}^{0} dt
=-\frac{E_{\mathbf{k}}^{0}-E_{\mathbf{k0}}^{0}}{e\mathbf{\cal E}}.
\end{equation}
So the $E_{\mathbf{k}}^{0}$ at finite fields should be used to 
replace the quantity at zero field to analyze the Zener tunneling.


\begin{references}
\bibitem{14} A longer but more rigorous derivation is based on the formula 
$d \left( \frac{1}{N_{\mathbf{k}}} \sum_{\mathbf{k}}E_{\mathbf{k}}^0 \right) 
/{d \mathbf{\cal E}}=\mathbf{\cal E}\cdot
d \mathbf{P} / d \mathbf{\cal E}$ where $E_{\mathbf{k}}^0=\langle u_{\mathbf{k}%
}|H_{0}|u_{\mathbf{k}}\rangle$, which is not given in this paper.
\end{references}
\end{document}